\documentclass[journal=jacsat,manuscript=article]{achemso}

\usepackage[version=3]{mhchem} 

\usepackage{mathtools}
\usepackage{amssymb}
\usepackage{physics}
\usepackage{color}
\usepackage{graphicx}
\usepackage{tabularx}
\usepackage{dcolumn} 
\usepackage{bm}      
\usepackage{array}   
\usepackage{amsfonts}
\usepackage{geometry}
\usepackage{caption}
\usepackage{subcaption}
\usepackage{url}
\usepackage{amsmath}
\usepackage[utf8]{inputenc}
\usepackage[T1]{fontenc}

\date{}

\author{Lijun Wang}
\affiliation{Center for Scientific Computing, Cavendish Laboratory, University of Cambridge, Cambridge}
\altaffiliation{Yusuf Hamied Department of Chemistry,University of Cambridge, Cambridge}
\email{lw755@cam.ac.uk}

\title{Sparse Full Configuration Interaction}

\begin{document}



\begin{abstract}
We propose an efficient deterministic method to calculate the full configuration interaction (FCI) ground state energy. This method leverages the sparseness of the Lanczos basis vectors that span the Krylov subspace associated with the Hamiltonian to rapidly calculate the lowest eigenvalue of the effective Hamiltonian formed in this basis. By iteratively performing the spanning and diagonalization steps, this method is capable of rapidly reaching chemical accuracy for a variety of strongly correlated molecules, including stretched N$_2$ and C$_2$, in merely several tens of iterations. The accuracy is systematically improved by increasing the number of Slater determinants included in a single Lanczos basis vector. To accelerate our algorithm, we implement parallelized sparse matrix-sparse vector multiplication, which features a novel hashing method that establishes a one-to-one correspondence between the set of full configuration Slater determinants and a set of consecutive integers. A near-linear acceleration was observed. 

\end{abstract}

\section{1. Introduction}

In order to predict the reactivity of molecules or the properties of quantum materials, one is typically faced with having to solve the quantum system of interest's Schr\"odinger Equation, what is thought to be an NP-hard problem that has led to the development of a multitude of approximate classical algorithms. Conceptually, the most straightforward way of solving the Schr\"odinger Equation is to identify all of the possible, physically-acceptable states of the system and diagonalize the system Hamiltonian formed in this basis of many-body states for its eigenvalues and eigenvectors, which can then be manipulated to obtain its quantum properties. This approach is known as the full configuration interaction (FCI) approach in the chemical literature\cite{fci_knowles_1984,fci_knowles_1989}, and the exact diagonalization (ED) approach in the physical literature\cite{ed_weisse_2008}. Since its computational cost scales combinatorially with the number of orbitals and electrons taken into account, FCI can only be applied to relatively small chemical systems whose wave functions can be expanded in no more than $10^{10}$ determinants \cite{ROSSI1999530,n2JCP}. 

Multiple methods have been developed to accurately approximate the FCI ground state energy at much lower cost, making it possible to study previously inaccessible quantum systems, especially the ones that involve strongly correlated electrons. The density matrix renormalization group (DMRG)\cite{dmrg_white_1992,ab_initio_dmrg_1999,spin_adapted_qcdmrg_chan_2012,dmrg_scf_marcel_2008,high_performance_qcdmrg_yanai_2012} method, from which a more generalized theory of tensor network states\cite{abdmrg_in_practice_chan_2015,efficient_mpo_reiher_2015,mps_mpo_abdmrg_white_2016,review_TNSinChem_szalay_2015,tree_tns_verstraete_2015,mps_large_site_verstraete_2022,complete_graph_state_2010,3leg_tns_versraete_2018} is derived, provides a compact parametrization of the wave function that only scales polynomially with the system size. Selected CI systematically truncates the exponentially large FCI spaces, only retaining the determinants that have the most significant contribution to the total energy. Beyond the early attempts\cite{Buenker1974,extarpolation_ci_Buenker_1975,cipsi_origin_Rancurel_1973,improved_cipsi_malrieu_1983,size_consistent_ci_perturb_malrieu_1994,determinant_ci_1988,multiref_pertb_ci_1997}, progresses in selected CI in the past decade feature heat-bath CI\cite{heat_bath_ci_tubman_2016,semi_stochatsic_hci_es_holmes_2017,semistochastic_hci_sandeep_2017}, adaptive CI\cite{adaptive_ci_evangelista_2016,adaptive_ci_excited_evangelista_2017}, iterative CI\cite{iterative_ci_liu_2016,iterative_ci_selection_liu_2020}, machine learning-based CI\cite{mlci},and adaptive sampling CI\cite{adaptive_sampling_ci_head-gordon_2016,asci_modern_approach_head-gordon_2020}. Extensive efforts have also been made in stochastically simulating the imaginary time evolution (ITE) $\hat{U}(\tau) = e^{-\hat{H}\tau}$. Auxiliary-field quantum Monte Carlo and its phase-less variant\cite{afqmc_koonin_1986,afqmc_gaussian_zhang_2006,phaseless_afqmc_zhang_2003} operate within spaces spanned by non-orthogonal determinants. In contrast, the full configuration interaction quantum Monte Carlo (FCIQMC)
method \cite{FCIQMC} expands the wave function with a collection of orthogonal determinant walkers, each of which carries either a positive or a negative sign. At discrete time, the dynamics of walkers is governed by $\ket{\Psi(\tau + \delta \tau)} = \ket{\Psi(\tau)} - \delta \tau \hat{H} \ket{\Psi(\tau)}$, and the calculation of the long-time limit $\lim_{\tau \rightarrow \infty} \ket{\Psi(\tau)}$ is enabled by sampling $\hat{H} \ket{\Psi(\tau)}$ at each time step. This algorithm thus employs an efficient representation of the wave function, and more importantly, circumvents the use of the fixed node approximation \cite{anderson1975,klein1976,fixed_node_ceperly_1984} that are required in other stochastic ground state algorithms. Multiple modifications to and reformulations of FCIQMC have subsequently been introduced, including initiator approximation\cite{iFCIQMC,unbiasing_iFCIQMC_alavi_2019}, semi-stochastic projection\cite{SPMC,semi_stocahstic_FCIQMC_blunt_2015}, Chebyshev projector expansion\cite{chebyshev_ime_evangelista_2016}, and fast randomized iteration\cite{fast_randomized_iteration_berkelbach_2019,improved_fri_berkelbach_2020}. Similar ideas have been applied to devise coupled cluster Monte Carlo\cite{stochatsic_CC_thom_2010,multiref_stochastic_cc_thom_2019,hybrid_ci_ccmc_thom_2023}. It has also been attempted to sample the intermediate states from the ITE trajectory to construct a basis that enables efficient calculations of isolated excited states, thermal, and spectral properties\cite{kryloc_projected_fciqmc_blunt_2015}.

As observed by Neufeld and Thom, an irksome problem found in FCIQMC is that the ``computational effort to reach equilibration can be very significant"\cite{quasinewton}. In practice, $10^5$ to $10^6$ iterations are required before the energy converges. \cite{FCIQMC,iFCIQMC,semistochasticfciqmc}. 
Such a behavior can be understood in terms of the projector's form. Numerically, the power method (PM) is used to approach the ground state in FCIQMC. Suppose $A$ is an operator, and it has eigenvalues $\abs{\lambda_0} > \abs{\lambda_1} >...> \abs{\lambda_N}$. The PM obtains the eigenvector that corresponds to $\lambda_0$ by iteratively applying $A$ onto a reasonably chosen trial vector. The convergence rate of PM is determined by $\abs{{\lambda_1} / {\lambda_0}}$ \cite{saad1992}. In the projector methods, usually $A = e^{-\delta \tau (\hat{H}-S) }$ (where $S$ is a parameter used for walker population control and $\delta \tau$ a small time step), and the PM's rate of convergence is $e^{-(E_1-E_0)\delta \tau} \approx 1 - (E_1-E_0)\delta \tau$, where $E_0$ and $E_1$ are the two lowest eigenvalues of $H$. Since $E_1-E_0$ is usually of order $10^{-1}$ a.u., and $\Delta \tau$ ranges from $10^{-3}$ to $10^{-4}$ a.u.\cite{FCIQMC,iFCIQMC}, the rate of convergence is near-unity in practice, and is arguably a partial reason that FCIQMC-like algorithms are usually costly. 


Although new techniques have been proposed to address this shortcoming \cite{quasinewton,fciqmc-preconditioning}, new solutions are worth exploring. 
Here, we propose a deterministic method to approximate the FCI ground state energy more efficiently. We call our method sparse full configuration interaction (SFCI). It is essentially a projector method that also employs ideas from selected CI. In contrast with the FCIQMC family of methods, SFCI uses a process resembling the Lanczos method to span the search space. It starts with a trial wave function $\ket{v_0}$, and generates a basis that consists of $N_L$ vectors $ \{ \ket{w_{i}} \}$ which span the Krylov subspace. However, unlike in the Lanczos method, each SFCI basis vector is truncated with respect to the magnitude of the coefficients, and only contains up to $N_{d}$ elements. $N_{d}$ and $N_L$ are both user-defined parameters. With this basis, an $N_L$-by-$N_L$ effective Hamiltonian $T$ is formed and diagonalized. The eigenvector corresponding to the lowest eigenvalue is then used to construct a new trial wave function for the next round of spanning and diagonalizing. This process is repeated until convergence is reached. 

Since we use sparse vectors to represent the Lanczos basis, the efficiency of the whole algorithm relies heavily on that of the sparse matrix-sparse vector multiplication (SpMSpVM). We thus propose and implement a parallel sparse matrix-sparse vector multiplication algorithm. It is based on the algorithm proposed by Azad and Buluc\cite{spmspv}, which uses a ``bucketing" technique to allow different cores to merge their intermediate vectors without communicating with each other. Unlike in this algorithm's original setting, in which non-zero elements of the sparse matrix are explicitly stored, the matrix elements in our application are generated on-the-fly. The absolute index (i.e., the position in the FCI vector) of any determinant is thus unknown, and so the bucketing is hindered. To fill this gap, we introduce an efficient indexing method that establishes a one-to-one correspondence between the FCI determinants and a set of consecutive integers [0,1,2,...N-1]. By combining these methods, substantial speedup is achieved. In various tests, our parallelized SpMSpVM implementation achieves speedups of around 12-fold on 32 cores with respect to 2 cores. 


This article consists of three parts. First, we introduce the theoretical underpinnings of the SFCI method in Section 2. We next discuss the implementation details of our parallelized sparse matrix-sparse vector multiplication algorithm in Section 3. Finally, in Section 4, we present the results obtained by SFCI on a range of strongly-correlated small molecules, and discuss possible improvements.


\section{2. Method}
\label{SFCI}
\subsection{2.1 Sparse Full Configuration Interaction (SFCI) Iterative Solver}

Proposed in 1950 \cite{Lanczos1950AnIM}, the Lanczos method has become a powerful tool for finding the extreme eigenvalues and eigenvectors of large sparse self-adjoint matrices. This has led to its widespread application in quantum physics and chemistry, which often seek to obtain ground state wave functions and energies. The Lanczos method is attractive because the number of iterations it requires for accurately approximating the ground state is much smaller than the size of the problem, i.e., the exponentially-scaling dimensionality of the Hilbert space, $\dim(\mathcal{H})$. To our best knowledge, convergence to the ground state is usually achieved within several tens of iterations. 

Here, we briefly outline the Lanczos method \cite{Koch2011}. Given a sparse Hamiltonian $H$ and a wave function $\ket{v_0}$, the $N+1$-dimensional Krylov subspace is defined as $\mathcal{K}_{N+1}(\ket{v_0},H) = \text{span} \{v_0, H \ket{v_0}, H^2 \ket{v_0},..., H^N \ket{v_0}\}$. The Lanczos method then constructs an orthonormal basis $\{\ket{u_i}\}$ in the Krylov subspace, and diagonalizes the effective Hamiltonian $T$ formed under this basis. In practice, the number of $\ket{u_i}$ required for a converged result (e.g., most commonly, the few eigenvalues that reside at either end of the spectrum) is usually much smaller than the dimension of the Hilbert space. 

Such an orthonormal basis is constructed as follows. Given a normalized trial wave function $\ket{v_0}$, let $\ket{u_0} = \ket{v_0}$. $\ket{u_1}$ can next be obtained via a Gram-Schmidt orthogonalization:  

\begin{equation}
\begin{aligned}
	\ket{u'_1} &= H\ket{u_0} - \bra{u_0} H \ket{u_0} \ket{u_0} \\
				& =  H\ket{u_0} - a_0 \ket{u_0}
\end{aligned}
\end{equation}
with 
\begin{equation}
\begin{aligned}
	\ket{u_1} &= \ket{u'_1} /\sqrt{\braket{u'_1}}.
\end{aligned}
\end{equation}
Similarly,
\begin{equation}
\begin{aligned}
	\ket{u'_2} &= H\ket{u_1}  - \bra{u_1} H \ket{u_1} \ket{u_1} - \bra{u_0} H \ket{u_1} \ket{u_0}\\
	& =  H\ket{u_1} - a_1 \ket{u_1} - b_1\ket{u_0} \\ 
	\ket{u_2} &= \ket{u'_2} /\sqrt{\braket{u'_2}} 
\end{aligned}
\end{equation}
Notice that $\ket{u_3}$ becomes independent of $\ket{u_0}$
\begin{equation}
\begin{aligned}
	\ket{u'_3} &= H\ket{u_2}  - \bra{u_2} H \ket{u_2} \ket{u_2} - \bra{u_1} H \ket{u_2} \ket{u_1} - \bra{u_0} H \ket{u_2} \ket{u_0}\\
	& =  H\ket{u_2} - a_2 \ket{u_2} - b_2\ket{u_1} - 0\\ 
	\ket{u_3} &= \ket{u'_3} /\sqrt{\braket{u'_3}} 
\end{aligned}
\end{equation}
since $\bra{u_0} H \ket{u_2}$ vanishes under the definition in Equations (1) and (3). This leads to a recursive formula to calculate the $i^{th}$ basis vector
\begin{equation}
\begin{aligned}
\ket{u'_i} &= H\ket{u_{i-1}}  - \bra{u_{i-1}} H \ket{u_{i-1}} \ket{u_{i-1}} - \bra{u_{i-2}} H \ket{u_{i-1}} \ket{u_{i-2}}\\
& =  H\ket{u_{i-1}} - a_{i-1} \ket{u_{i-1}} - b_{i-1}\ket{u_{i-2}} \\ 
\ket{u_{i}} &= \ket{u'_i} /\sqrt{\braket{u'_i}} 
\end{aligned}
\end{equation}
with
\begin{equation}
\begin{aligned}
a_i &= \bra{u_i} H \ket{u_i} \\
b_i &= \bra{u_{i-1}} H \ket{u_i} = \sqrt{\braket{u_{i}}}.
\end{aligned}
\end{equation}
In other words, $T$ is \textit{tri-diagonal} under the basis $\{\ket{u_i}\}$
\begin{equation}
    T_{ij} = 
    \begin{cases}
        \bra{u_i} H \ket{u_j}, & |i-j| \leq 1 \\
        0, &\text{otherwise}
    \end{cases}
\end{equation}
With $\lambda_0$ being $T$'s smallest eigenvalue and $x_0 = (c_1,c_2,...,c_{L})$ its corresponding eigenvector, the approximated ground state wave function is
\begin{equation}
    \ket{\Psi_0} = \sum c_i \ket{u_i}
\end{equation}
and the ground state energy is approximated by 
\begin{equation}
    E_0 = \frac{\bra{\Psi_0} H \ket{\Psi_0}}{\braket{\Psi_0}}.
\end{equation}


In our method, the construction of the basis $\{\ket{w_i}\}$ differs slightly from that employed by the Lanczos method: each basis vector $\ket{w_i}$ only contains $N_d$ of the most important determinants. That is, one first obtains
\begin{equation}
\begin{aligned}
    \ket{u_i} & = H\ket{w_{i-1}}  - \bra{w_{i-1}} H \ket{w_{i-1}} \ket{w_{i-1}} - \bra{w_{i-2}} H \ket{w_{i-1}} \ket{w_{i-2}}\\
    & = \sum c_j \ket{D_j},\ \abs{c_1} > \abs{c_2} > ... > \abs{c_n}
\end{aligned}
\end{equation}
in accordance with Equation (5),  
and truncates the result with respect to the amplitudes
\begin{equation}
\begin{aligned}
	\ket{w'_i} &= \sum_{j}^{N_d} c_j \ket{D_j}\\
	\ket{w_i} &= \ket{w'_i} / \sqrt{\braket{w'_i}}.
\end{aligned}
\end{equation}
This restriction avoids the exponential growth in the number of determinants after each application of $H$. Notice that since all $\ket{w_i}$ are truncated, they are slightly off from being mutually orthogonal. Therefore, the problem of interest becomes a generalized eigenvalue problem, $Tx = \lambda S x$,
where $S_{ij} = \braket{w_i}{w_j}$. In principle, one could construct a large set of basis vectors and diagonalize $T$ only once. But in practice, we enforce the process to restart after $N_L$ basis vectors $\{\ket{w_1}, \ket{w_2},..., \ket{w_{N_L}}\}$ are collected. It goes as follows: diagonalizing the $n^{th}$ effective Hamiltonian yields an estimated ground state $\ket{\Psi^{(n)}_0}$, which is next truncated down to $N_d$ determinants and used as the the initial basis vector $\ket{w^{(n+1)}_0}$ to form $(n+1)^{th}$ $T$. The restarting approach is also widely adopted \cite{YSaad,templates,saad1992} in the application of the Lanczos method. It helps to circumvent the loss of orthogonality of basis as the iteration number increases, which could otherwise introduce spurious solutions that are not physical\cite{Koch2011,spurious}, as well as to reduce the peak memeory usage. 
Finally, our algorithm halts when $\bra{\Psi^{(n-1)}_0} H \ket{\Psi^{(n-1)}_0} - \bra{\Psi^{(n)}_0} H \ket{\Psi^{(n)}_0}$ falls below a threshold $\epsilon$.

\subsection{2.2 Sparse Matrix Sparse Vector Multiplication in Parallel}
\label{sparse}

Since all wave functions in SFCI are represented by sparse vectors, an efficient sparse matrix-sparse vector multiplication is desirable. However, although parallel sparse matrix-dense vector multiplication (SpMVM) has been implemented in various libraries, shifting to the case of sparse matrix-sparse vector multiplication (SpMSpVM) requires extra care. We now demonstrate a novel SpMSpVM algorithm that particularly suits problems in determinant spaces. We restrict our discussion to the way the algorithm would function on a shared-memory machine in this article. As will be shown later, migrating this algorithm to a distributed memory system is straightforward.

\begin{figure}
	\centering
	\includegraphics[width=1.0\linewidth]{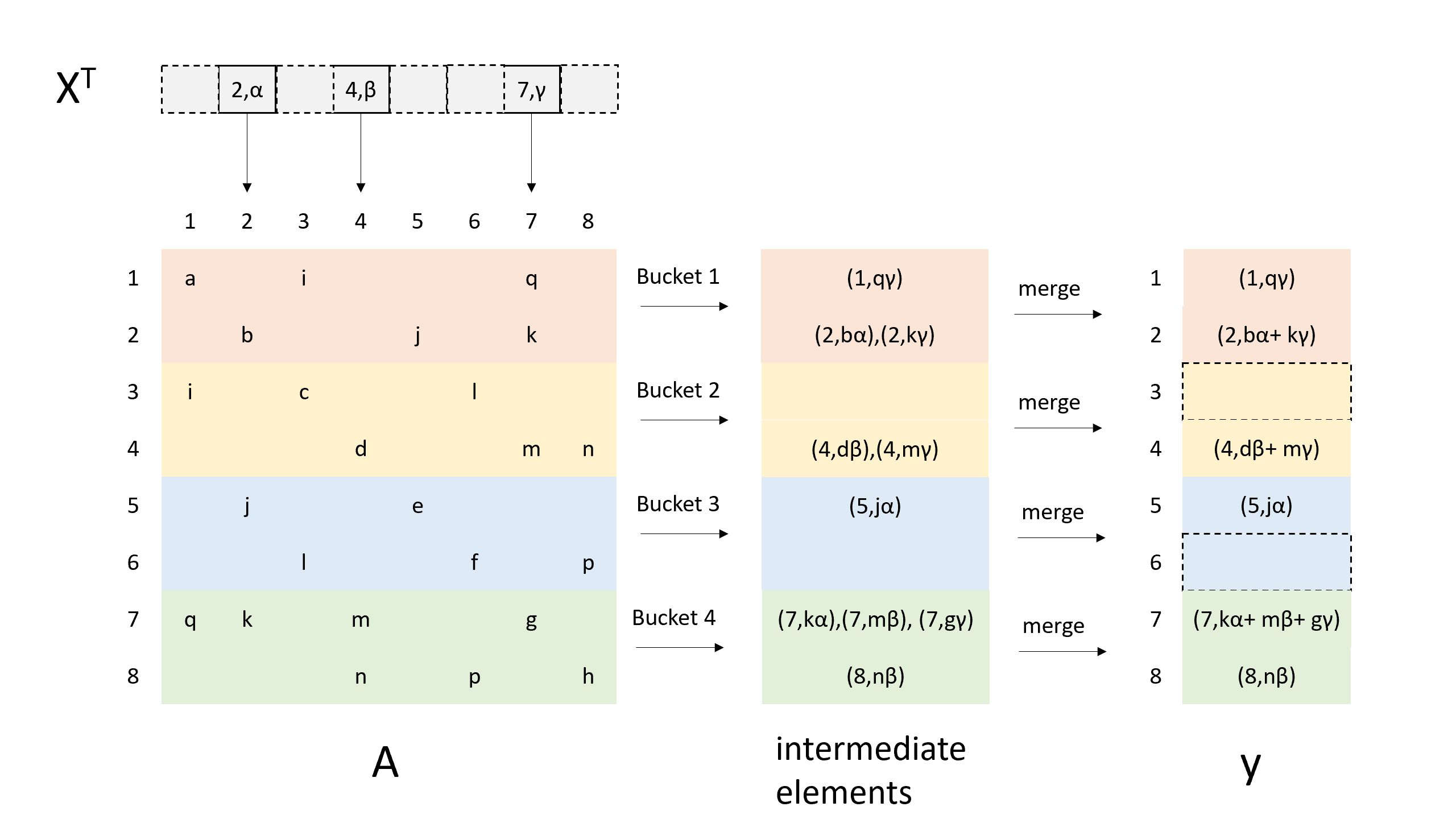}
	\caption{Demonstration of the bucketing in a SpMSpVM calculation. Dashed blocks in $x^T$ and $y$ stands for zero elements. For clarity, the sparse matrix $A$ is shown as a square rather than as 3-tuples. Buckets are labeled with different colors. Notice that, without the bucketing, all intermediate elements have to be merged by 1 core, but 4 cores can now work on the merging simultaneously with bucketing.}
   \label{fig_1}
\end{figure}

We implement the parallel SpMSpVM algorithm based on the ``bucketing strategy" proposed by Azad and Buluc \cite{spmspv}. The advantage of such a design is that it enables all available computational resources to simultaneously work on the merging part of the SpMSpVM calculation. Suppose the problem of interest is $Ax=y$, where $A$ is a sparse matrix and $x,y$ are sparse vectors. We represent a sparse vector as a set of 2-tuples $(i,\alpha_i)$, where $\alpha_i$ is the non-zero element and $i$ its index within the array (see Figure \ref{fig_1} for a depiction). A sparse matrix is instead a set of 3-tuples $(i,j,c_{ij})$, where $i,j$ are the row and column indices, and $c_{ij}$ the non-zero element. There are two major steps in a SpMSpVM calculation. First, for each $(i,\alpha_i)$ in $x$, all of its connected product 2-tuples $\{(k,\alpha_ic_{ki})\}$ are generated by going through all 3-tuples with matching indices $\{(k,i,c_{ki})\}$ in $A$. We call those 2-tuples generated from $x$ \textit{intermediate elements}. Second, the set of intermediate elements is merged, i.e., all 2-tuples with identical indices $(i, \alpha^1_i)$, $(i,\alpha^2_i)$, $(i,\alpha^3_i)$, etc., are merged into a single 2-tuple $(i,\sum_j \alpha^j_i)$. The merged set of intermediate elements is just the resultant sparse vector $y$. Parallelizing step 1 is straightforward, as one could assign each available CPU core with a segment of $x$ and generate the intermediate elements simultaneously. It is less obvious how step 2 can be executed in parallel, but this is where the bucketing becomes advantageous. As shown in Figure \ref{fig_1}, the intermediate elements are distributed across buckets based on their indices. Each bucket can be merged by different CPU cores without communication between them. Finally, the merged buckets are concatenated to form the result $y$. 

In practice, when the input vector $x$ has a large number of elements, $N_{T}$ cores will independently calculate their own segments to generate all of the intermediate elements. To avoid a data race, each core is assigned an independent copy of the buckets. When all intermediate elements are generated, these local buckets are brought together to form the final set of buckets. A sort-and-merge scheme is used to merge each bucket: first, all elements in the bucket are sorted with respect to the determinant, so that the coefficients of a single determinant are brought together. A scan from the beginning to the end will next merge these coefficients. We use the introspective sort as the sorting method, whose average and worst case complexity are both $O(n\log n)$ \cite{introsort}. The complexity of the scanning step is the typical $O(n)$. 

In FCI, the matrix elements are calculated before the optimization starts. First, the complete set of FCI determinants is generated, so that any determinant's index within a dense FCI vector can be determined. One may create for later convenience a hash table $h$ with the determinant as its key and index its value. Next, for all determinants $\ket{D_i}$, with $i$ being $\ket{D_i}$'s index in the FCI vector, all its connected determinants $\ket{D_j}$ are calculated. $j$ can be easily found using the previously built hash table $h$. All 3-tuples ($i$, $j$, $\bra{D_i} H \ket{D_j}$) are finally stored in the sparse matrix data structure, which will be used later in sparse matrix-dense vector multiplications. In contrast, since it is impossible to explicitly store all matrix elements for large systems, matrix elements must be generated on the fly. That is, given any determinant $\ket{D}$, all of its connected determinants $\ket{D_\alpha}$, as well as the corresponding matrix elements $ \{\bra{D_\alpha} H \ket{D}\}$, are calculated. There is, however, no way to know a determinant's index when the system becomes large, since it is impossible to explicitly generate an FCI vector. The bucketing thus breaks down for large systems. 

To tackle this problem, we introduce a novel hashing method to enable bucketing under matrix-free SpMSpVM in determinant spaces. Essentially, such a method establishes a one-to-one correspondence between a set of FCI determinants $\{\ket{D_1},\ket{D_2},...,\ket{D_N}\}$, and a set of consecutive integers $\{0,1,2,...,N-1\}$. Suppose $\ket{D} = \ket{\chi_1\chi_2 ... \chi_M}$, where $\chi_i$ is the orbital index occupied by the ith electron and ranges from [1,$N_{orb}$]. The hash value of $\ket{D}$ is calculated using the following formula: 

\begin{equation}
\begin{aligned}
	h(\ket{D}) = \sum_{i=1}^M \binom{\chi_{i}-1 }{i}
\end{aligned}
\end{equation}
in which all $\binom{n}{k}$ values with $n$ smaller than $k$ are set to 0 for convenience. 

Such a formula is derived from two observations. First, given a properly chosen``initial" determinant, all determinants in the FCI space can be enumerated \textit{recursively}. That is, given any bit string that represents a determinant, its unique successor can be found by a function $g$. The only exception is the ``final" determinant, which has no successor. To that end, we define the initial determinant in any system as 11..1100..000, in which all 1-bits are flushed to the left. Consequently, the final determinant can be defined as 00..001..11, in which all 1-bits are flushed to the right. We then define function $g$ as follows. For any determinant written in bit string form, $g$:

\begin{enumerate}
    \item[] Scan from the left all 1-bits in the bit string. If a 1-bit has a 0-bit as its right neighbor, flush all 1-bits on its left to the left, swap these two bits, and return the new bit string as the successor; if none of the 1-bits have a 0-bit as a right neighbor, the final determinant has been reached. 
\end{enumerate} 
In the table below, we use a system with 6 orbitals and 3 electrons to demonstrate how to enumerate the entire FCI space using $g$ starting with the initial determinant. 
\begin{table*}[h]
	\centering
	\caption{Generation of the FCI space of a (6o,3e) system via $g$. }
	$\begin{array}{ c }
	
	111000 \xrightarrow{\textit{g}} 110100 \xrightarrow{\textit{g}} 101100 \xrightarrow{\textit{g}} 011100 \xrightarrow{\textit{g}} 110010 \\ 
	
	\xrightarrow{\textit{g}} 101010 \xrightarrow{\textit{g}}  011010\xrightarrow{\textit{g}} 100110\xrightarrow{\textit{g}} 010110\xrightarrow{\textit{g}} 001110 \\
	
	\xrightarrow{\textit{g}}110001 	\xrightarrow{\textit{g}}101001	\xrightarrow{\textit{g}} 011001	\xrightarrow{\textit{g}} 100101	\xrightarrow{\textit{g}} 010101 \\
	
    \xrightarrow{\textit{g}} 001101 \xrightarrow{\textit{g}}100011 \xrightarrow{\textit{g}}010011 \xrightarrow{\textit{g}}001011 \xrightarrow{\textit{g}}000111 

	\end{array}$
\end{table*}
We call the above enumerating process a ``g-chain" for later convenience. Now, it is obvious that the distance from the initial determinant, i.e. how many times $g$ has been applied, uniquely defines any determinant in the FCI space, and the value of distance can therefore be used as the index of the corresponding determinant. 

Second, to calculate the distance, observe that it can be further decomposed into the contributions made by each 1-bit. We use the following example to demonstrate how to calculate these contributions. Consider the determinant in the bit-string form 101010. The rightmost 1-bit occupies orbital 5, meaning that before arriving at the current determinant, one must, at least, pass through all determinants in the g-chain with 3 electrons occupying the first 4 orbitals from the initial determinant. We thus take the contribution from this 1-bit as $\binom{4}{3} = 4$. Next, the second 1-bit occupies orbital 3. To reach the configuration with the third electron occupying orbital 5 and second electron orbital 3, one has to pass through all determinants with the first 2 electrons occupying the first 2 orbitals and the third electron in orbital 5, whose total number is $\binom{2}{2} = 1$. Finally, the first electron is not shifted from its initial orbital, yielding a contribution $\binom{0}{1} = 0$. The distance, of 101010 is therefore $4+1+0=5$, which we take as its hash value. This can be verified in the g-chain shown above, in which the initial determinant has hash value 0. A complete list of the hash values of this system is given in Table 2.  

\begin{table*}[h]
	\centering
	\caption{Hash values of all determinants with 6 orbitals and 3 electrons  }
	$\begin{array}{ c c }
    \hline
    
	h(\ket{111000}) =  \binom{0}{1}+\binom{1}{2}+\binom{2}{3} = 0 & 
	h(\ket{110001}) =  \binom{0}{1}+\binom{1}{2}+\binom{5}{3} = 10 \\

	h(\ket{110100}) =  \binom{0}{1}+\binom{1}{2}+\binom{3}{3} = 1 & 
	h(\ket{101001}) =  \binom{0}{1}+\binom{2}{2}+\binom{5}{3} = 11 \\

	h(\ket{101100}) =  \binom{0}{1}+\binom{2}{2}+\binom{3}{3} = 2 & 
	h(\ket{011001}) =  \binom{1}{1}+\binom{2}{2}+\binom{5}{3} = 12\\

	h(\ket{011100}) =  \binom{1}{1}+\binom{2}{2}+\binom{3}{3} = 3 &
	h(\ket{100101}) =  \binom{0}{1}+\binom{3}{2}+\binom{5}{3} = 13	\\

	h(\ket{110010}) =  \binom{0}{1}+\binom{1}{2}+\binom{4}{3} = 4 & 
	h(\ket{010101}) =  \binom{1}{1}+\binom{3}{2}+\binom{5}{3} = 14\\

	h(\ket{101010}) =  \binom{0}{1}+\binom{2}{2}+\binom{4}{3} = 5 & 
	h(\ket{001101}) =  \binom{2}{1}+\binom{3}{2}+\binom{5}{3} = 15	\\

	h(\ket{011010}) =  \binom{1}{1}+\binom{2}{2}+\binom{4}{3} = 6 &
	h(\ket{100011}) =  \binom{0}{1}+\binom{4}{2}+\binom{5}{3} = 16\\

	h(\ket{100110}) =  \binom{0}{1}+\binom{3}{2}+\binom{4}{3} = 7 & 
	h(\ket{010011}) =  \binom{1}{1}+\binom{4}{2}+\binom{5}{3} = 17\\

	h(\ket{010110}) =  \binom{1}{1}+\binom{3}{2}+\binom{4}{3} = 8 & 
	h(\ket{001011}) =  \binom{2}{1}+\binom{4}{2}+\binom{5}{3} = 18\\

	h(\ket{001110}) =  \binom{2}{1}+\binom{3}{2}+\binom{4}{3} = 9 & 
	h(\ket{000111}) =  \binom{3}{1}+\binom{4}{2}+\binom{5}{3} = 19\\
	\hline
	
	\end{array}$
\end{table*}

For any determinant $\ket{D}$ that needs to be bucketed, its hash value $h(\ket{D})$ is used to determine the bucket to which it belongs. Given the pre-defined bucket size $N_B$(i.e. the number of indices it contains), the bucket index of $\ket{D}$ is the floor of $\frac{h(\ket{D})}{N_B}$, i.e., $\lfloor \frac{h(\ket{D})}{N_B}\rfloor$. 

\section{3. Computational Details}

In this work, we consider molecular Hamiltonians that contain up to double excitations. We use MolPro \cite{molpro} to calculate the restricted Hartree-Fock (RHF) determinant as well as the associated electron-electron integrals. The RHF determinant is used as the trial wave function. Matrix elements $\bra{D_i} H \ket{D_j}$ in the SFCI calculation are generated via our own implementation in C++. We use a 64-bit unsigned integer in which a 1-bit corresponds to an occupied spin orbital, while a 0-bit an unoccupied spin orbital to represent determinants. Though at this point such a representation can only accommodate determinants with no more than 64 spin orbitals, this representation can be generalized by, for instance, using more than one integer to represent a single determinant.  Additionally, in all of our calculations, we compute the bucketing indices only with respect to the spin-up section of the bit-string. For closed-shell systems, this can reduce the time cost by a factor of 2. Finally, all of our numerical experiments are carried out on a single node containing two 2.80GHz Intel Xeon Gold 6242 CPUs, with 32 cores in total. All cores are used in the experiments unless otherwise specified.  

In practice, we introduce an extra parameter to accelerate the calculation of  $H \ket{u_{i}}$, which is the most expensive part of the whole SFCI routine. As described in the previous sections, the merged intermediate vectors from different buckets in $H \ket{u_{i}}$ are concatenated into one vector $\ket{v}$, and the most important $N_d$ determinants are picked out to form $\ket{u_{i+1}}$. This requires sorting the whole $\ket{v}$ vector. However, sorting $\ket{v}$ can be expensive, since its length can be a thousand times larger than that of $\ket{u_{i}}$ in practice. We therefore introduce $\epsilon$ as the threshold for incorporating the determinants into $\ket{v}$. No determinant in the intermediate vectors with coefficient amplitude below $\epsilon$ is incorporated into $\ket{v}$. Such a truncation in principle influences the values of all $T_{ij},j<i$, but in the experiments, we find that it has no visible effect on the accuracy of SFCI. It is also noteworthy that, in some cases, an $\epsilon$ that is too small can lead to non-physical solutions and complicate convergence. In all tests below, we use $\epsilon = 5\times10^{-5}$.

The source code that implements the above can be accessed at \cite{repo}.

\section{4. Results}
\label{results}
\subsection{4.1 Sparse-Matrix Sparse-Vector Acceleration}

We first demonstrate the acceleration achieved by our SpMSpVM algorithm. Figure 2 shows the acceleration ratios of a single SpMSpVM calculation with different numbers of cores for different systems. We use 2 cores as the baseline since bucketing is meaningless on 1 core. The tests manifest clearly that our SpMSpVM implementation is able to achieve a similar, near linear speedup for different Hamiltonians and input vector sizes. 

\begin{figure*}[!ht]
	\centering
	\begin{subfigure}{.5\textwidth}
      \centering
      \includegraphics[width=1.0\linewidth]{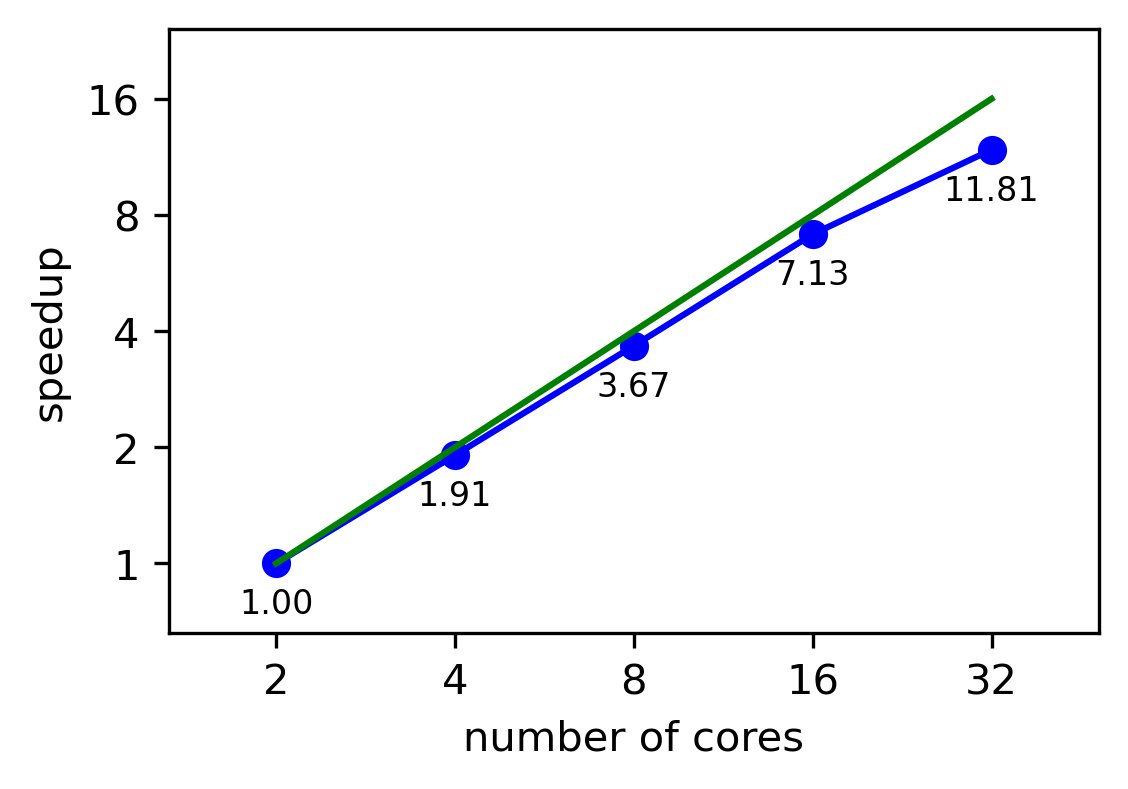}
      \caption{C$_2$, cc-pVDZ, $N_d = 4\times10^5$}
    \end{subfigure}%
    \begin{subfigure}{.5\textwidth}
      \centering
      \includegraphics[width=1.0\linewidth]{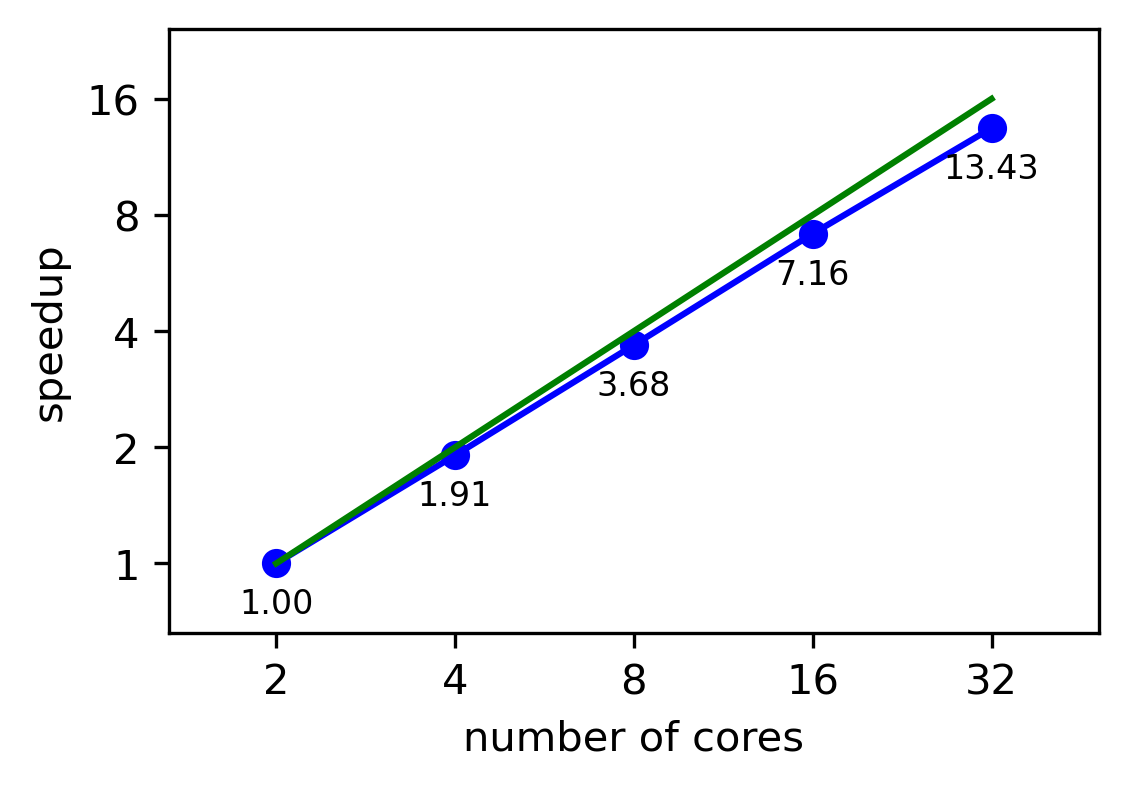}
      \caption{C$_2$, cc-pVDZ, $N_d = 8\times10^5$ dets}
    \end{subfigure} %
    \begin{subfigure}{.5\textwidth}
      \centering
      \includegraphics[width=1.0\linewidth]{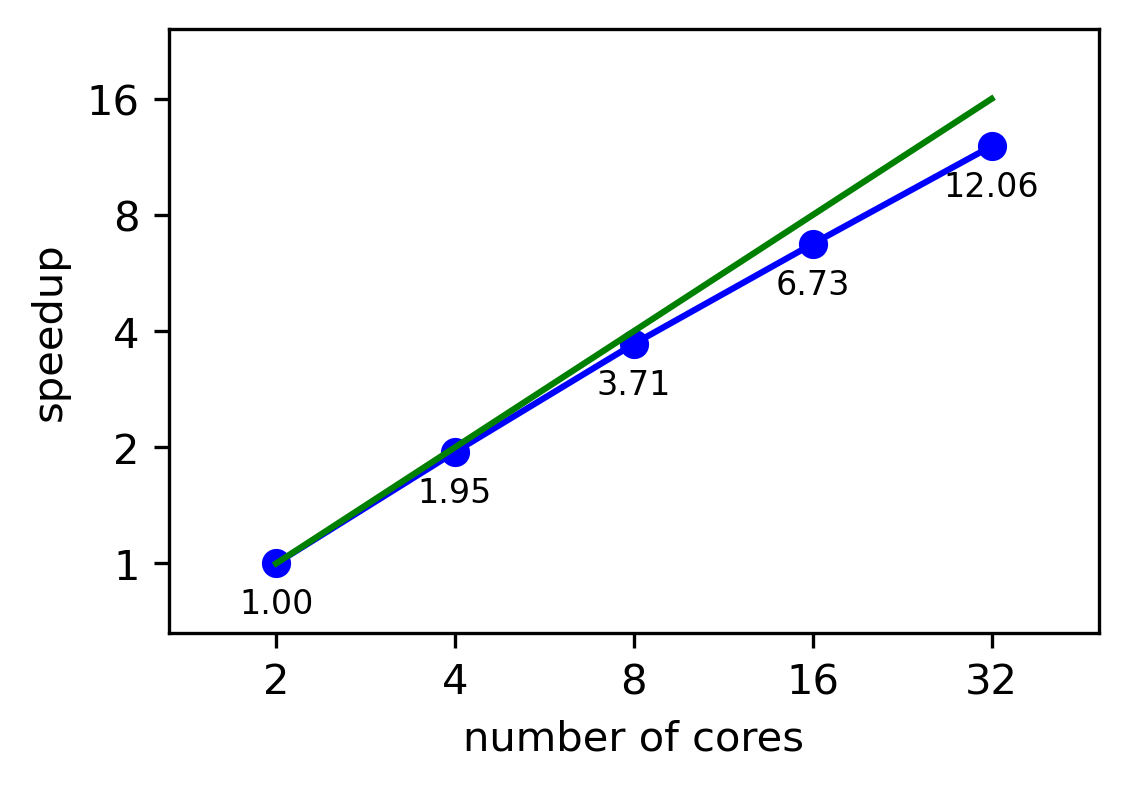}
      \caption{N$_2$, cc-pVDZ, $N_d = 4\times10^5$ dets}
    \end{subfigure}
    
	\caption{Relative parallel SpMSpVM speedup for different molecular systems. The blue dots represent the acceleration ratio relative to 2 cores, while the green line is the ideal linear scaling with respect to the core number. The three cases that have been plotted are: C$_2$ using the cc-pVDZ basis with input vector of size $N_d = 4\times10^5$, C$_2$ using the cc-pVDZ basis with input vector of size $N_d = 8\times10^5$, and N$_2$ using the cc-pVDZ basis, with input vector $N_d = 4\times10^5$.}
	
\end{figure*}

\subsection{4.2 Convergence of SFCI Algorithm with Respect to the Number of Determinants and Basis Size}

We further test the convergence of SFCI, as well as how parameters $N_d$ and $N_L$ impact that convergence. Table 3 provides the SFCI results for C$_2$ and N$_2$, multiply-bonded molecules with known FCI benchmark energies. C$_2$ is analyzed at its equilibrium geometry (bond distance of 1.27273 \r{A}), while N$_2$ is analyzed in both its equilibrium (bond distance of 1.09397 \r{A}) and stretched (bond distance of 2.22180 \r{A}) geometry. The above geometries are from Ref \cite{FCIQMC}. In all calculations, $N_L$ (i.e., the dimensionality of the effective Hamiltonian) is fixed at 10. It can be seen that 1 mHa accuracy is easily obtained in all cases within spaces that are merely 1\% as large as the FCI spaces.. For C$_2$, the best SFCI energy is 0.13 mHa above the FCI result. For N$_2$, the best results are 0.29 mHa and 0.43 mHa above the FCI results for the equilibrium and stretched geometries, respectively. To give an idea of the computation time, C$_2$ with $N_d=4\times10^5$ takes 10 minutes, while N$_2$ in equilibrium with $N_d=6\times 10^5$ takes 14 minutes. It is also clear that as $N_d$ increases, the error of the converged result decreases in all three cases. Also, the size of the resulting wave function, $\ket{\phi_0}$, is roughly proportional to $N_d$, indicating that increasing $N_d$ can lead to a larger solution space. 

SFCI may also be observed to converge rapidly with respect to the iteration number $N_I$. We define the iteration number as how many times SpMSpVM is applied before convergence is achieved, which is simply $N_L$ times the number of effective Hamiltonians that have been diagonalized. As shown in Table 3, the $N_I$ required for convergence never exceeds 100 in all of our experiments, indicating that SFCI has successfully inherited the Lanczos method's advantage of rapid convergence.

\begin{center}
    \small
	\captionof{table}{SFCI energies as a function of the number of determinants and iterations. C$_2$ is studied in its  equilibrium geometry while N$_2$ is studied in both its equilibrium and stretched geometries in the cc-pVDZ basis set with a frozen core. This yields a (26o,8e) configuration for C$_2$, and a (26,10e) configuration for N$_2$. The size of the FCI spaces in the $D_{2h}$ point group are $2.79\times 10^7$ and $5.41\times10^8$, respectively. The FCI result for C$_2$ is obtained via MolPro, while those for N$_2$ in its equilibrium and stretched geometries are obtained from References \cite{n2JCP} and \cite{N2-fci-2004}, respectively. Energies are in units of Hartrees.} 
	
	\begin{tabular}{c c c c c c c} 
		\hline
		System & $N_d / 10^5$ & size of $\ket{\phi_0}$/$10^5$ & $N_I$ & $E_{SFCI}$  & $E_{FCI}$ & $E_{FCI}$ - $E_{SFCI}$/mHa \\ 
		\hline\hline

	   C$_2$ & 2 & 5.8 & 50 & -75.729413 & -75.729852 & 0.439 \\ 

		 & 4 &10.4 & 50& -75.729664 &     & 0.188 \\ 
		 
		 & 6 &12.0 & 40& -75.729667 & & 0.184\\
		 
		 & 8 &13.7 & 50& -75.729696 & & 0.155 \\
		 
		 & 10 &18.2 & 50& -75.729717 & & 0.135 \\ 
		\hline
		N$_2$ equilibrium& 2 & 5.5 & 40& -109.275647  & -109.276527 & 0.88 \\
		
	         & 4 & 11.3 & 40& -109.276097  &  & 0.430 \\
		
		     & 6 & 15.9 & 40& -109.276234  &  & 0.292 \\
		     
		     & 8 & 14.8 & 30& -109.276235 & & 0.292 \\
		     & 10 & 17.8 & 30& -109.276251 & & 0.276  \\ 
		\hline
		
		N$_2$ stretched& 2 &6.4 & 90& -108.96454  & -108.96695 & 2.409 \\
		
	         & 4 & 12.6 & 80& -108.96576  &  & 1.189 \\
		
		     & 6 & 19.0 &80& -108.96622  &  & 0.729 \\
		     
		     & 8 & 24.6 & 80& -108.96644 & & 0.509 \\
		     
		     & 10 & 24.2 & 80 & -108.96651 & & 0.43 \\
		     
		\hline 
	\end{tabular}
\end{center}

We observe that, in comparison to $N_d$, $N_L$ has a much less visible impact on the accuracy of the converged results. In the applications of the Lanczos method, a larger $N_L$ means more degrees of freedom for variation, resulting in faster convergence \cite{Koch2011}. However, we observe that neither the speed of convergence nor the SFCI algorithm's accuracy is significantly changed by varying $N_L$. Table 4 shows the converged energies and iteration numbers for different $N_L$ for the C$_2$ cc-pVDZ system. Notice that different $N_L$ only yield $10^{-2}$ mHa differences in the converged energies, and small differences in the number of iterations needed for convergence. This suggests that using a relatively small $N_L$ is not only economical, but reasonably accurate. 

\begin{center}
    \small
	\captionof{table}{ SFCI results of C$_2$ cc-pVDZ with different $N_L$ with $N_d$ is fixed to $4\times 10^5$. In contrast to one may expect, increasing $N_L$ benefits neither the algorithm's accuracy nor its speed of convergence.} 
	
	\begin{tabular}{c c c} 
		\hline
		$N_L$ & $E_{SFCI}$  &  $N_I$\\ 
		\hline\hline
        5 & -75.729646 & 50 \\
        10 & -75.729683 & 50 \\
        15 & -75.729689 & 75 \\
        20 & -75.729665 & 80 \\
		\hline
	\end{tabular}
\end{center}

\subsection{4.3 Quantum Chemical Benchmarks}

We finally test SFCI on several more challenging systems. Table 2 shows the SFCI results for systems for which FCI results are absent, as well as existing results obtained by other methods. In all cases, $N_L$ = 10, except for F$_2$, for which we use $N_L$ = 5. For Ne, CO, and C$_2$ with all of its electrons fully correlated, SFCI results are all within 0.5 mHa of the existing DMRG or FCIQMC results. On the other hand, F$_2$ stands as an example that is overly challenging for FCI calculations. For F$_2$ in its equilibrium geometry with all of its electrons correlated in the cc-pVDZ basis, its FCI space size reaches $5.96 \times 10^{12}$ under the $D_{2h}$ point group. With $N_d = 6 \times 10^5$, its SFCI energy is -199.102153 Ha, which is 1 mHa lower than the CCSD(t) energy of $E_{CCSD(t)}$ = -199.101154 Ha. The size of the resulting wave function is $1.24\times10^6$ determinants. In comparison, under a slightly different geometry, the ASCI method yields $E_{ASCI}$ = -199.101763 Ha with $10^6$ determinants and $E_{CCSD(t)}$ = -199.101481 Ha, a discrepancy of 0.28 mHa \cite{asci_modern_approach_head-gordon_2020}. However, the bare ASCI result is significantly improved by the second order Epstein-Nesbet correction, which yields an energy that is 1.864 mHa lower than the CCSD(t) result \cite{asci_modern_approach_head-gordon_2020}. This suggests that combining perturbative corrections with SFCI may further improve the latter's accuracy, a point for further future exploration. The longest computation among these examples is F$_2$ with $N_d = 6\times10^5$, which took 85 minutes. This further attests to the rapid convergence of SFCI. 

\begin{center}
    \small
    \captionof{table}
    {SFCI energies relative to those obtained from other methods. The geometries used for the diatomic molecules are: CO (r = 1.1448 \AA), C$_2$ (r= 1.24253\AA), and F$_2$(r = 1.412 \AA). For Ne, C$_2$, and F$_2$, all electrons (AE) are correlated, while for CO, the core electrons are frozen. The resultant configurations are : Ne(23o,10e), CO(26o,10e), C$_2$(28o,12e), and F$_2$(28o,18e). The $D_{2h}$ point group is used for all species presented except for CO, which is given in the $C_{2v}$ point group. The resultant sizes of the FCI spacces for these systems are: Ne ($1.4 \times 10^8$ determinants), CO($1.08\times 10^9$ determinants), C$_2$($1.77\times10^{10}$ determinants), and F$_2$ ($5.96\times10^{12}$ determinants). The FCIQMC result for CO is obtained from ref \cite{FCIQMC}. The DMRG result for Ne is obtained from \cite{kryloc_projected_fciqmc_blunt_2015}, and for C$_2$ from \cite{abinitiodmrg}. M in the table denotes the bond dimension used for the DMRG calculations. The CCSD(t) result for F$_2$ is calculated via MolPro.}

	\begin{tabular}{c c c c c c c} 
		\hline
		System & $N_{d}$ / $10^5 $ &$E_{SFCI}$   & $E_{DMRG}$ & $E_{FCIQMC}$ & $E_{CCSD(t)}$ & $N_I$\\ 
		\hline\hline

		Ne, AE  & 2 & -128.711461  &-128.71147 (M=500)& & &30\\ 
        \hline
		CO & 4 &-113.055552  & &-113.05644   & &40\\
         & 8 &-113.056045   & &  & & 50\\ 
          & 12 &-113.056066   & &  & & 30\\ 
        \hline
		C$_2$, AE & 4 &-75.731555   & -75.731958 (M=4000)& & &80 \\
	           & 8 &-75.731715   &  & & & 100\\
        \hline
		F$_2$, AE & 2 & -199.101305 & & & -199.101154 & 70 \\ 
		       & 4 & -199.1016715 & & &   & 80\\ 
		       & 6 & -199.102153 & & &  & 80 \\ 
		\hline
		
	\end{tabular}
\end{center}

We now briefly compare SFCI and Krylov-projected FCIQMC (KP-FCIQMC)\cite{kryloc_projected_fciqmc_blunt_2015}, which appears similar at first glance. Besides being a deterministic method, SFCI differs from KP-FCIQMC in both its projection approach as well as the basis chosen. In KP-FCIQMC, an FCIQMC-style projector $ \hat{P} = 1-\delta\tau(H-S)$ is used to define the Krylov subspace $\mathcal{K}$, and the basis for constructing the effective $H$ is of the form $\{ \hat{P}^n \ket{\psi_0} \}$, which is far from being orthogonal. But in SFCI, $H$ is used as in the Lanczos method to define $\mathcal{K}$, and the computational basis is designed to be orthogonal, though the truncation process introduces slight nonorthogonality in practice. Regarding the ground state calculation, this setting arguably makes SFCI more efficient. Also, SFCI and KP-FCIQMC adopt slightly different techniques for curtailing the increasing linear dependence of the basis vectors that appear during late stages of the algorithm. As shown above, SFCI restricts the number of basis vectors to a relatively small number and restarts its spanning and diagonalization multiple times; in contrast, KP-FCIQMC samples from the Krylov vectors with linearly increasing distances between selections, so that the vectors in the later stage of the algorithm, which have increasingly larger overlap with the previous ones, are less likely to be selected. 

Improving upon this algorithm by parallelizing SFCI over multiple compute nodes is straightforward, and we briefly outline the strategy here. Suppose N nodes are collaboratively carrying out a SpMSpVM calculation on the input wave function $\ket{x}$. Each node computes a segment of $\ket{x}$ simultaneously, and let $\ket{y_i}$ denote the resultant sparse vector from node i. Now, all $\ket{y_i}$ have to be merged and truncated into one single sparse vector. However, it is possible that a single node's memory is insufficient to accommodate all $\ket{y_i}$. An alternative approach is to merge $\ket{y_i}$ via a reduction tree. The reduction continues until only one active node remains. At each layer, the inter-node communications and the merging-and-truncation steps of all pairs take place simultaneously. The total time cost of the reduction will be $\log(N)$ times the time cost of a single layer. The calculation of the effective Hamiltonian $T$ and and the overlap matrix $S$ is finally carried out on the master node. 

We now discuss a few possibilities for future development. We notice that some recent advances in the selected CI method may also further accelerate SFCI. For example, since the current version of SFCI generates all single and double excitations of the reference determinants, incorporating heat-bath CI's efficient selection method\cite{heat_bath_ci_tubman_2016} will significantly reduced the time consumed in sorting and merging. Semi-stochastic projection \cite{SPMC,semistochasticfciqmc} may also lead to further improvements in efficiency. It is furthermore possible to find low-lying excited state energies via SFCI. In theory, the only modification required, is to properly project out all previously found eigenvectors in $H_{\text{eff}}$ when solving for the $i^{th}$ excited state\cite{GoluVanl96}. 

\section{5. Conclusions}

We have proposed and implemented an efficient deterministic method to approximate the FCI energy with modest computation resources. Compared to  FCIQMC-style methods, SFCI constructs a Lanczos-like basis to search for the ground state energy, thus requiring far fewer iterations to converge. We have shown that SFCI is capable of obtaining highly accurate ground state energies for various strongly correlated systems with FCI spaces of up to $5.96\times10^{12}$ determinants. Our novel indexing function as well as our implementation of the parallel sparse matrix and sparse vector multiplication have enabled us to achieve the above results in a matter of an hour. We believe that, in combination with various other techniques, the efficiency and the accuracy of SFCI can be further improved, making it possible to study a wide variety of challenging problems in quantum many body physics and chemistry.

\section{Data Availability}

All codes, scripts, and data needed to reproduce the results in this paper are available online at: 
\url{https://github.com/wlj89/SFCI}.

\section{Acknowledgements}
The author is greatly indebted to Brenda Rubenstein for support and guidance, and thanks Alex Thom for helpful discussions. This research was conducted using computational resources and services at the Center for Computation and Visualization, Brown University.

\bibliography{ref}

\providecommand{\latin}[1]{#1}
\makeatletter
\providecommand{\doi}
  {\begingroup\let\do\@makeother\dospecials
  \catcode`\{=1 \catcode`\}=2 \doi@aux}
\providecommand{\doi@aux}[1]{\endgroup\texttt{#1}}
\makeatother
\providecommand*\mcitethebibliography{\thebibliography}
\csname @ifundefined\endcsname{endmcitethebibliography}  {\let\endmcitethebibliography\endthebibliography}{}
\begin{mcitethebibliography}{70}
\providecommand*\natexlab[1]{#1}
\providecommand*\mciteSetBstSublistMode[1]{}
\providecommand*\mciteSetBstMaxWidthForm[2]{}
\providecommand*\mciteBstWouldAddEndPuncttrue
  {\def\EndOfBibitem{\unskip.}}
\providecommand*\mciteBstWouldAddEndPunctfalse
  {\let\EndOfBibitem\relax}
\providecommand*\mciteSetBstMidEndSepPunct[3]{}
\providecommand*\mciteSetBstSublistLabelBeginEnd[3]{}
\providecommand*\EndOfBibitem{}
\mciteSetBstSublistMode{f}
\mciteSetBstMaxWidthForm{subitem}{(\alph{mcitesubitemcount})}
\mciteSetBstSublistLabelBeginEnd
  {\mcitemaxwidthsubitemform\space}
  {\relax}
  {\relax}

\bibitem[Knowles and Handy(1984)Knowles, and Handy]{fci_knowles_1984}
Knowles,~P.; Handy,~N. A new determinant-based full configuration interaction method. \emph{Chemical Physics Letters} \textbf{1984}, \emph{111}, 315--321\relax
\mciteBstWouldAddEndPuncttrue
\mciteSetBstMidEndSepPunct{\mcitedefaultmidpunct}
{\mcitedefaultendpunct}{\mcitedefaultseppunct}\relax
\EndOfBibitem
\bibitem[Knowles and Handy(1989)Knowles, and Handy]{fci_knowles_1989}
Knowles,~P.~J.; Handy,~N.~C. A determinant based full configuration interaction program. \emph{Computer Physics Communications} \textbf{1989}, \emph{54}, 75--83\relax
\mciteBstWouldAddEndPuncttrue
\mciteSetBstMidEndSepPunct{\mcitedefaultmidpunct}
{\mcitedefaultendpunct}{\mcitedefaultseppunct}\relax
\EndOfBibitem
\bibitem[Wei{\ss}e and Fehske(2008)Wei{\ss}e, and Fehske]{ed_weisse_2008}
Wei{\ss}e,~A.; Fehske,~H. In \emph{Computational Many-Particle Physics}; Fehske,~H., Schneider,~R., Wei{\ss}e,~A., Eds.; Springer Berlin Heidelberg: Berlin, Heidelberg, 2008; pp 529--544\relax
\mciteBstWouldAddEndPuncttrue
\mciteSetBstMidEndSepPunct{\mcitedefaultmidpunct}
{\mcitedefaultendpunct}{\mcitedefaultseppunct}\relax
\EndOfBibitem
\bibitem[Rossi \latin{et~al.}(1999)Rossi, Bendazzoli, Evangelisti, and Maynau]{ROSSI1999530}
Rossi,~E.; Bendazzoli,~G.~L.; Evangelisti,~S.; Maynau,~D. A full-configuration benchmark for the N2 molecule. \emph{Chemical Physics Letters} \textbf{1999}, \emph{310}, 530--536\relax
\mciteBstWouldAddEndPuncttrue
\mciteSetBstMidEndSepPunct{\mcitedefaultmidpunct}
{\mcitedefaultendpunct}{\mcitedefaultseppunct}\relax
\EndOfBibitem
\bibitem[Gan \latin{et~al.}(2006)Gan, Grant, Harrison, and Dixon]{n2JCP}
Gan,~Z.; Grant,~D.~J.; Harrison,~R.~J.; Dixon,~D.~A. The lowest energy states of the group-IIIA–group-VA heteronuclear diatomics: BN, BP, AlN, and AlP from full configuration interaction calculations. \emph{The Journal of Chemical Physics} \textbf{2006}, \emph{125}, 124311\relax
\mciteBstWouldAddEndPuncttrue
\mciteSetBstMidEndSepPunct{\mcitedefaultmidpunct}
{\mcitedefaultendpunct}{\mcitedefaultseppunct}\relax
\EndOfBibitem
\bibitem[White(1992)]{dmrg_white_1992}
White,~S.~R. Density matrix formulation for quantum renormalization groups. \emph{Phys. Rev. Lett.} \textbf{1992}, \emph{69}, 2863--2866\relax
\mciteBstWouldAddEndPuncttrue
\mciteSetBstMidEndSepPunct{\mcitedefaultmidpunct}
{\mcitedefaultendpunct}{\mcitedefaultseppunct}\relax
\EndOfBibitem
\bibitem[White and Martin(1999)White, and Martin]{ab_initio_dmrg_1999}
White,~S.~R.; Martin,~R.~L. {Ab initio quantum chemistry using the density matrix renormalization group}. \emph{The Journal of Chemical Physics} \textbf{1999}, \emph{110}, 4127--4130\relax
\mciteBstWouldAddEndPuncttrue
\mciteSetBstMidEndSepPunct{\mcitedefaultmidpunct}
{\mcitedefaultendpunct}{\mcitedefaultseppunct}\relax
\EndOfBibitem
\bibitem[Sharma and Chan(2012)Sharma, and Chan]{spin_adapted_qcdmrg_chan_2012}
Sharma,~S.; Chan,~G. K.-L. {Spin-adapted density matrix renormalization group algorithms for quantum chemistry}. \emph{The Journal of Chemical Physics} \textbf{2012}, \emph{136}, 124121\relax
\mciteBstWouldAddEndPuncttrue
\mciteSetBstMidEndSepPunct{\mcitedefaultmidpunct}
{\mcitedefaultendpunct}{\mcitedefaultseppunct}\relax
\EndOfBibitem
\bibitem[Zgid and Nooijen(2008)Zgid, and Nooijen]{dmrg_scf_marcel_2008}
Zgid,~D.; Nooijen,~M. {The density matrix renormalization group self-consistent field method: Orbital optimization with the density matrix renormalization group method in the active space}. \emph{The Journal of Chemical Physics} \textbf{2008}, \emph{128}, 144116\relax
\mciteBstWouldAddEndPuncttrue
\mciteSetBstMidEndSepPunct{\mcitedefaultmidpunct}
{\mcitedefaultendpunct}{\mcitedefaultseppunct}\relax
\EndOfBibitem
\bibitem[Kurashige and Yanai(2009)Kurashige, and Yanai]{high_performance_qcdmrg_yanai_2012}
Kurashige,~Y.; Yanai,~T. {High-performance ab initio density matrix renormalization group method: Applicability to large-scale multireference problems for metal compounds}. \emph{The Journal of Chemical Physics} \textbf{2009}, \emph{130}, 234114\relax
\mciteBstWouldAddEndPuncttrue
\mciteSetBstMidEndSepPunct{\mcitedefaultmidpunct}
{\mcitedefaultendpunct}{\mcitedefaultseppunct}\relax
\EndOfBibitem
\bibitem[Olivares-Amaya \latin{et~al.}(2015)Olivares-Amaya, Hu, Nakatani, Sharma, Yang, and Chan]{abdmrg_in_practice_chan_2015}
Olivares-Amaya,~R.; Hu,~W.; Nakatani,~N.; Sharma,~S.; Yang,~J.; Chan,~G. K.-L. {The ab-initio density matrix renormalization group in practice}. \emph{The Journal of Chemical Physics} \textbf{2015}, \emph{142}, 034102\relax
\mciteBstWouldAddEndPuncttrue
\mciteSetBstMidEndSepPunct{\mcitedefaultmidpunct}
{\mcitedefaultendpunct}{\mcitedefaultseppunct}\relax
\EndOfBibitem
\bibitem[Keller \latin{et~al.}(2015)Keller, Dolfi, Troyer, and Reiher]{efficient_mpo_reiher_2015}
Keller,~S.; Dolfi,~M.; Troyer,~M.; Reiher,~M. {An efficient matrix product operator representation of the quantum chemical Hamiltonian}. \emph{The Journal of Chemical Physics} \textbf{2015}, \emph{143}, 244118\relax
\mciteBstWouldAddEndPuncttrue
\mciteSetBstMidEndSepPunct{\mcitedefaultmidpunct}
{\mcitedefaultendpunct}{\mcitedefaultseppunct}\relax
\EndOfBibitem
\bibitem[Chan \latin{et~al.}(2016)Chan, Keselman, Nakatani, Li, and White]{mps_mpo_abdmrg_white_2016}
Chan,~G. K.-L.; Keselman,~A.; Nakatani,~N.; Li,~Z.; White,~S.~R. {Matrix product operators, matrix product states, and ab initio density matrix renormalization group algorithms}. \emph{The Journal of Chemical Physics} \textbf{2016}, \emph{145}, 014102\relax
\mciteBstWouldAddEndPuncttrue
\mciteSetBstMidEndSepPunct{\mcitedefaultmidpunct}
{\mcitedefaultendpunct}{\mcitedefaultseppunct}\relax
\EndOfBibitem
\bibitem[Szalay \latin{et~al.}(2015)Szalay, Pfeffer, Murg, Barcza, Verstraete, Schneider, and Legeza]{review_TNSinChem_szalay_2015}
Szalay,~S.; Pfeffer,~M.; Murg,~V.; Barcza,~G.; Verstraete,~F.; Schneider,~R.; Legeza,~O. Tensor product methods and entanglement optimization for ab initio quantum chemistry. \emph{International Journal of Quantum Chemistry} \textbf{2015}, \emph{115}, 1342--1391\relax
\mciteBstWouldAddEndPuncttrue
\mciteSetBstMidEndSepPunct{\mcitedefaultmidpunct}
{\mcitedefaultendpunct}{\mcitedefaultseppunct}\relax
\EndOfBibitem
\bibitem[Murg \latin{et~al.}(2015)Murg, Verstraete, Schneider, Nagy, and Legeza]{tree_tns_verstraete_2015}
Murg,~V.; Verstraete,~F.; Schneider,~R.; Nagy,~P.~R.; Legeza,~O. Tree Tensor Network State with Variable Tensor Order: An Efficient Multireference Method for Strongly Correlated Systems. \emph{Journal of Chemical Theory and Computation} \textbf{2015}, \emph{11}, 1027--1036, PMID: 25844072\relax
\mciteBstWouldAddEndPuncttrue
\mciteSetBstMidEndSepPunct{\mcitedefaultmidpunct}
{\mcitedefaultendpunct}{\mcitedefaultseppunct}\relax
\EndOfBibitem
\bibitem[Larsson \latin{et~al.}(2022)Larsson, Zhai, Gunst, and Chan]{mps_large_site_verstraete_2022}
Larsson,~H.~R.; Zhai,~H.; Gunst,~K.; Chan,~G. K.-L. Matrix Product States with Large Sites. \emph{Journal of Chemical Theory and Computation} \textbf{2022}, \emph{18}, 749--762\relax
\mciteBstWouldAddEndPuncttrue
\mciteSetBstMidEndSepPunct{\mcitedefaultmidpunct}
{\mcitedefaultendpunct}{\mcitedefaultseppunct}\relax
\EndOfBibitem
\bibitem[Marti \latin{et~al.}(2010)Marti, Bauer, Reiher, Troyer, and Verstraete]{complete_graph_state_2010}
Marti,~K.~H.; Bauer,~B.; Reiher,~M.; Troyer,~M.; Verstraete,~F. Complete-graph tensor network states: a new fermionic wave function ansatz for molecules. \emph{New Journal of Physics} \textbf{2010}, \emph{12}, 103008\relax
\mciteBstWouldAddEndPuncttrue
\mciteSetBstMidEndSepPunct{\mcitedefaultmidpunct}
{\mcitedefaultendpunct}{\mcitedefaultseppunct}\relax
\EndOfBibitem
\bibitem[Gunst \latin{et~al.}(2018)Gunst, Verstraete, Wouters, Legeza, and Van~Neck]{3leg_tns_versraete_2018}
Gunst,~K.; Verstraete,~F.; Wouters,~S.; Legeza,~O.; Van~Neck,~D. T3NS: Three-Legged Tree Tensor Network States. \emph{Journal of Chemical Theory and Computation} \textbf{2018}, \emph{14}, 2026--2033\relax
\mciteBstWouldAddEndPuncttrue
\mciteSetBstMidEndSepPunct{\mcitedefaultmidpunct}
{\mcitedefaultendpunct}{\mcitedefaultseppunct}\relax
\EndOfBibitem
\bibitem[Buenker and Peyerimhoff(1974)Buenker, and Peyerimhoff]{Buenker1974}
Buenker,~R.~J.; Peyerimhoff,~S.~D. Individualized configuration selection in CI calculations with subsequent energy extrapolation. \emph{Theoretica chimica acta} \textbf{1974}, \emph{35}, 33--58\relax
\mciteBstWouldAddEndPuncttrue
\mciteSetBstMidEndSepPunct{\mcitedefaultmidpunct}
{\mcitedefaultendpunct}{\mcitedefaultseppunct}\relax
\EndOfBibitem
\bibitem[Buenker and Peyerimhoff(1975)Buenker, and Peyerimhoff]{extarpolation_ci_Buenker_1975}
Buenker,~R.~J.; Peyerimhoff,~S.~D. Energy extrapolation in CI calculations. \emph{Theoretica chimica acta} \textbf{1975}, \emph{39}, 217--228\relax
\mciteBstWouldAddEndPuncttrue
\mciteSetBstMidEndSepPunct{\mcitedefaultmidpunct}
{\mcitedefaultendpunct}{\mcitedefaultseppunct}\relax
\EndOfBibitem
\bibitem[Huron \latin{et~al.}(2003)Huron, Malrieu, and Rancurel]{cipsi_origin_Rancurel_1973}
Huron,~B.; Malrieu,~J.~P.; Rancurel,~P. {Iterative perturbation calculations of ground and excited state energies from multiconfigurational zeroth‐order wavefunctions}. \emph{The Journal of Chemical Physics} \textbf{2003}, \emph{58}, 5745--5759\relax
\mciteBstWouldAddEndPuncttrue
\mciteSetBstMidEndSepPunct{\mcitedefaultmidpunct}
{\mcitedefaultendpunct}{\mcitedefaultseppunct}\relax
\EndOfBibitem
\bibitem[Evangelisti \latin{et~al.}(1983)Evangelisti, Daudey, and Malrieu]{improved_cipsi_malrieu_1983}
Evangelisti,~S.; Daudey,~J.-P.; Malrieu,~J.-P. Convergence of an improved CIPSI algorithm. \emph{Chemical Physics} \textbf{1983}, \emph{75}, 91--102\relax
\mciteBstWouldAddEndPuncttrue
\mciteSetBstMidEndSepPunct{\mcitedefaultmidpunct}
{\mcitedefaultendpunct}{\mcitedefaultseppunct}\relax
\EndOfBibitem
\bibitem[Meller \latin{et~al.}(1994)Meller, Heully, and Malrieu]{size_consistent_ci_perturb_malrieu_1994}
Meller,~J.; Heully,~J.; Malrieu,~J. Size-consistent self-consistent combination of selected CI and perturbation theory. \emph{Chemical Physics Letters} \textbf{1994}, \emph{218}, 276--282\relax
\mciteBstWouldAddEndPuncttrue
\mciteSetBstMidEndSepPunct{\mcitedefaultmidpunct}
{\mcitedefaultendpunct}{\mcitedefaultseppunct}\relax
\EndOfBibitem
\bibitem[Olsen \latin{et~al.}(1988)Olsen, Roos, Jo/rgensen, and Jensen]{determinant_ci_1988}
Olsen,~J.; Roos,~B.~O.; Jo/rgensen,~P.; Jensen,~H. J.~A. {Determinant based configuration interaction algorithms for complete and restricted configuration interaction spaces}. \emph{The Journal of Chemical Physics} \textbf{1988}, \emph{89}, 2185--2192\relax
\mciteBstWouldAddEndPuncttrue
\mciteSetBstMidEndSepPunct{\mcitedefaultmidpunct}
{\mcitedefaultendpunct}{\mcitedefaultseppunct}\relax
\EndOfBibitem
\bibitem[Angeli \latin{et~al.}(1997)Angeli, Cimiraglia, Persico, and Toniolo]{multiref_pertb_ci_1997}
Angeli,~C.; Cimiraglia,~R.; Persico,~M.; Toniolo,~A. Multireference perturbation CI I. Extrapolation procedures with CAS or selected zero-order spaces. \emph{Theoretical Chemistry Accounts} \textbf{1997}, \emph{98}, 57--63\relax
\mciteBstWouldAddEndPuncttrue
\mciteSetBstMidEndSepPunct{\mcitedefaultmidpunct}
{\mcitedefaultendpunct}{\mcitedefaultseppunct}\relax
\EndOfBibitem
\bibitem[Holmes \latin{et~al.}(2016)Holmes, Tubman, and Umrigar]{heat_bath_ci_tubman_2016}
Holmes,~A.~A.; Tubman,~N.~M.; Umrigar,~C.~J. Heat-Bath Configuration Interaction: An Efficient Selected Configuration Interaction Algorithm Inspired by Heat-Bath Sampling. \emph{Journal of Chemical Theory and Computation} \textbf{2016}, \emph{12}, 3674--3680, PMID: 27428771\relax
\mciteBstWouldAddEndPuncttrue
\mciteSetBstMidEndSepPunct{\mcitedefaultmidpunct}
{\mcitedefaultendpunct}{\mcitedefaultseppunct}\relax
\EndOfBibitem
\bibitem[Holmes \latin{et~al.}(2017)Holmes, Umrigar, and Sharma]{semi_stochatsic_hci_es_holmes_2017}
Holmes,~A.~A.; Umrigar,~C.~J.; Sharma,~S. {Excited states using semistochastic heat-bath configuration interaction}. \emph{The Journal of Chemical Physics} \textbf{2017}, \emph{147}, 164111\relax
\mciteBstWouldAddEndPuncttrue
\mciteSetBstMidEndSepPunct{\mcitedefaultmidpunct}
{\mcitedefaultendpunct}{\mcitedefaultseppunct}\relax
\EndOfBibitem
\bibitem[Sharma \latin{et~al.}(2017)Sharma, Holmes, Jeanmairet, Alavi, and Umrigar]{semistochastic_hci_sandeep_2017}
Sharma,~S.; Holmes,~A.~A.; Jeanmairet,~G.; Alavi,~A.; Umrigar,~C.~J. Semistochastic Heat-Bath Configuration Interaction Method: Selected Configuration Interaction with Semistochastic Perturbation Theory. \emph{Journal of Chemical Theory and Computation} \textbf{2017}, \emph{13}, 1595--1604\relax
\mciteBstWouldAddEndPuncttrue
\mciteSetBstMidEndSepPunct{\mcitedefaultmidpunct}
{\mcitedefaultendpunct}{\mcitedefaultseppunct}\relax
\EndOfBibitem
\bibitem[Schriber and Evangelista(2016)Schriber, and Evangelista]{adaptive_ci_evangelista_2016}
Schriber,~J.~B.; Evangelista,~F.~A. Communication: An adaptive configuration interaction approach for strongly correlated electrons with tunable accuracy. \emph{The Journal of Chemical Physics} \textbf{2016}, \emph{144}, 161106\relax
\mciteBstWouldAddEndPuncttrue
\mciteSetBstMidEndSepPunct{\mcitedefaultmidpunct}
{\mcitedefaultendpunct}{\mcitedefaultseppunct}\relax
\EndOfBibitem
\bibitem[Schriber and Evangelista(2017)Schriber, and Evangelista]{adaptive_ci_excited_evangelista_2017}
Schriber,~J.~B.; Evangelista,~F.~A. Adaptive Configuration Interaction for Computing Challenging Electronic Excited States with Tunable Accuracy. \emph{Journal of Chemical Theory and Computation} \textbf{2017}, \emph{13}, 5354--5366, PMID: 28892621\relax
\mciteBstWouldAddEndPuncttrue
\mciteSetBstMidEndSepPunct{\mcitedefaultmidpunct}
{\mcitedefaultendpunct}{\mcitedefaultseppunct}\relax
\EndOfBibitem
\bibitem[Liu and Hoffmann(2016)Liu, and Hoffmann]{iterative_ci_liu_2016}
Liu,~W.; Hoffmann,~M.~R. iCI: Iterative CI toward full CI. \emph{Journal of Chemical Theory and Computation} \textbf{2016}, \emph{12}, 1169--1178, PMID: 26765279\relax
\mciteBstWouldAddEndPuncttrue
\mciteSetBstMidEndSepPunct{\mcitedefaultmidpunct}
{\mcitedefaultendpunct}{\mcitedefaultseppunct}\relax
\EndOfBibitem
\bibitem[Zhang \latin{et~al.}(2020)Zhang, Liu, and Hoffmann]{iterative_ci_selection_liu_2020}
Zhang,~N.; Liu,~W.; Hoffmann,~M.~R. Iterative Configuration Interaction with Selection. \emph{Journal of Chemical Theory and Computation} \textbf{2020}, \emph{16}, 2296--2316\relax
\mciteBstWouldAddEndPuncttrue
\mciteSetBstMidEndSepPunct{\mcitedefaultmidpunct}
{\mcitedefaultendpunct}{\mcitedefaultseppunct}\relax
\EndOfBibitem
\bibitem[Coe(2018)]{mlci}
Coe,~J.~P. Machine Learning Configuration Interaction. \emph{Journal of Chemical Theory and Computation} \textbf{2018}, \emph{14}, 5739--5749, PMID: 30285426\relax
\mciteBstWouldAddEndPuncttrue
\mciteSetBstMidEndSepPunct{\mcitedefaultmidpunct}
{\mcitedefaultendpunct}{\mcitedefaultseppunct}\relax
\EndOfBibitem
\bibitem[Tubman \latin{et~al.}(2016)Tubman, Lee, Takeshita, Head-Gordon, and Whaley]{adaptive_sampling_ci_head-gordon_2016}
Tubman,~N.~M.; Lee,~J.; Takeshita,~T.~Y.; Head-Gordon,~M.; Whaley,~K.~B. A deterministic alternative to the full configuration interaction quantum Monte Carlo method. \emph{The Journal of Chemical Physics} \textbf{2016}, \emph{145}, 044112\relax
\mciteBstWouldAddEndPuncttrue
\mciteSetBstMidEndSepPunct{\mcitedefaultmidpunct}
{\mcitedefaultendpunct}{\mcitedefaultseppunct}\relax
\EndOfBibitem
\bibitem[Tubman \latin{et~al.}(2020)Tubman, Freeman, Levine, Hait, Head-Gordon, and Whaley]{asci_modern_approach_head-gordon_2020}
Tubman,~N.~M.; Freeman,~C.~D.; Levine,~D.~S.; Hait,~D.; Head-Gordon,~M.; Whaley,~K.~B. Modern Approaches to Exact Diagonalization and Selected Configuration Interaction with the Adaptive Sampling CI Method. \emph{Journal of Chemical Theory and Computation} \textbf{2020}, \emph{16}, 2139--2159\relax
\mciteBstWouldAddEndPuncttrue
\mciteSetBstMidEndSepPunct{\mcitedefaultmidpunct}
{\mcitedefaultendpunct}{\mcitedefaultseppunct}\relax
\EndOfBibitem
\bibitem[Sugiyama and Koonin(1986)Sugiyama, and Koonin]{afqmc_koonin_1986}
Sugiyama,~G.; Koonin,~S. Auxiliary field Monte-Carlo for quantum many-body ground states. \emph{Annals of Physics} \textbf{1986}, \emph{168}, 1--26\relax
\mciteBstWouldAddEndPuncttrue
\mciteSetBstMidEndSepPunct{\mcitedefaultmidpunct}
{\mcitedefaultendpunct}{\mcitedefaultseppunct}\relax
\EndOfBibitem
\bibitem[Al-Saidi \latin{et~al.}(2006)Al-Saidi, Zhang, and Krakauer]{afqmc_gaussian_zhang_2006}
Al-Saidi,~W.~A.; Zhang,~S.; Krakauer,~H. {Auxiliary-field quantum Monte Carlo calculations of molecular systems with a Gaussian basis}. \emph{The Journal of Chemical Physics} \textbf{2006}, \emph{124}, 224101\relax
\mciteBstWouldAddEndPuncttrue
\mciteSetBstMidEndSepPunct{\mcitedefaultmidpunct}
{\mcitedefaultendpunct}{\mcitedefaultseppunct}\relax
\EndOfBibitem
\bibitem[Zhang and Krakauer(2003)Zhang, and Krakauer]{phaseless_afqmc_zhang_2003}
Zhang,~S.; Krakauer,~H. Quantum Monte Carlo Method using Phase-Free Random Walks with Slater Determinants. \emph{Phys. Rev. Lett.} \textbf{2003}, \emph{90}, 136401\relax
\mciteBstWouldAddEndPuncttrue
\mciteSetBstMidEndSepPunct{\mcitedefaultmidpunct}
{\mcitedefaultendpunct}{\mcitedefaultseppunct}\relax
\EndOfBibitem
\bibitem[Booth \latin{et~al.}(2009)Booth, Thom, and Alavi]{FCIQMC}
Booth,~G.~H.; Thom,~A. J.~W.; Alavi,~A. Fermion Monte Carlo without fixed nodes: A game of life, death, and annihilation in Slater determinant space. \emph{The Journal of Chemical Physics} \textbf{2009}, \emph{131}, 054106\relax
\mciteBstWouldAddEndPuncttrue
\mciteSetBstMidEndSepPunct{\mcitedefaultmidpunct}
{\mcitedefaultendpunct}{\mcitedefaultseppunct}\relax
\EndOfBibitem
\bibitem[Anderson(1975)]{anderson1975}
Anderson,~J.~B. A random‐walk simulation of the Schrödinger equation: H+3. \emph{The Journal of Chemical Physics} \textbf{1975}, \emph{63}, 1499--1503\relax
\mciteBstWouldAddEndPuncttrue
\mciteSetBstMidEndSepPunct{\mcitedefaultmidpunct}
{\mcitedefaultendpunct}{\mcitedefaultseppunct}\relax
\EndOfBibitem
\bibitem[Klein and Pickett(1976)Klein, and Pickett]{klein1976}
Klein,~D.~J.; Pickett,~H.~M. Nodal hypersurfaces and Anderson’s random‐walk simulation of the Schrödinger equation. \emph{The Journal of Chemical Physics} \textbf{1976}, \emph{64}, 4811--4812\relax
\mciteBstWouldAddEndPuncttrue
\mciteSetBstMidEndSepPunct{\mcitedefaultmidpunct}
{\mcitedefaultendpunct}{\mcitedefaultseppunct}\relax
\EndOfBibitem
\bibitem[Ceperley and Alder(1984)Ceperley, and Alder]{fixed_node_ceperly_1984}
Ceperley,~D.~M.; Alder,~B.~J. {Quantum Monte Carlo for molecules: Green’s function and nodal release}. \emph{The Journal of Chemical Physics} \textbf{1984}, \emph{81}, 5833--5844\relax
\mciteBstWouldAddEndPuncttrue
\mciteSetBstMidEndSepPunct{\mcitedefaultmidpunct}
{\mcitedefaultendpunct}{\mcitedefaultseppunct}\relax
\EndOfBibitem
\bibitem[Cleland \latin{et~al.}(2010)Cleland, Booth, and Alavi]{iFCIQMC}
Cleland,~D.; Booth,~G.~H.; Alavi,~A. Communications: Survival of the fittest: Accelerating convergence in full configuration-interaction quantum Monte Carlo. \emph{The Journal of Chemical Physics} \textbf{2010}, \emph{132}, 041103\relax
\mciteBstWouldAddEndPuncttrue
\mciteSetBstMidEndSepPunct{\mcitedefaultmidpunct}
{\mcitedefaultendpunct}{\mcitedefaultseppunct}\relax
\EndOfBibitem
\bibitem[Ghanem \latin{et~al.}(2019)Ghanem, Lozovoi, and Alavi]{unbiasing_iFCIQMC_alavi_2019}
Ghanem,~K.; Lozovoi,~A.~Y.; Alavi,~A. {Unbiasing the initiator approximation in full configuration interaction quantum Monte Carlo}. \emph{The Journal of Chemical Physics} \textbf{2019}, \emph{151}, 224108\relax
\mciteBstWouldAddEndPuncttrue
\mciteSetBstMidEndSepPunct{\mcitedefaultmidpunct}
{\mcitedefaultendpunct}{\mcitedefaultseppunct}\relax
\EndOfBibitem
\bibitem[Petruzielo \latin{et~al.}(2012)Petruzielo, Holmes, Changlani, Nightingale, and Umrigar]{SPMC}
Petruzielo,~F.~R.; Holmes,~A.~A.; Changlani,~H.~J.; Nightingale,~M.~P.; Umrigar,~C.~J. Semistochastic Projector Monte Carlo Method. \emph{Phys. Rev. Lett.} \textbf{2012}, \emph{109}, 230201\relax
\mciteBstWouldAddEndPuncttrue
\mciteSetBstMidEndSepPunct{\mcitedefaultmidpunct}
{\mcitedefaultendpunct}{\mcitedefaultseppunct}\relax
\EndOfBibitem
\bibitem[Blunt \latin{et~al.}(2015)Blunt, Smart, Kersten, Spencer, Booth, and Alavi]{semi_stocahstic_FCIQMC_blunt_2015}
Blunt,~N.~S.; Smart,~S.~D.; Kersten,~J. A.~F.; Spencer,~J.~S.; Booth,~G.~H.; Alavi,~A. {Semi-stochastic full configuration interaction quantum Monte Carlo: Developments and application}. \emph{The Journal of Chemical Physics} \textbf{2015}, \emph{142}, 184107\relax
\mciteBstWouldAddEndPuncttrue
\mciteSetBstMidEndSepPunct{\mcitedefaultmidpunct}
{\mcitedefaultendpunct}{\mcitedefaultseppunct}\relax
\EndOfBibitem
\bibitem[Zhang and Evangelista(2016)Zhang, and Evangelista]{chebyshev_ime_evangelista_2016}
Zhang,~T.; Evangelista,~F.~A. A Deterministic Projector Configuration Interaction Approach for the Ground State of Quantum Many-Body Systems. \emph{Journal of Chemical Theory and Computation} \textbf{2016}, \emph{12}, 4326--4337, PMID: 27464301\relax
\mciteBstWouldAddEndPuncttrue
\mciteSetBstMidEndSepPunct{\mcitedefaultmidpunct}
{\mcitedefaultendpunct}{\mcitedefaultseppunct}\relax
\EndOfBibitem
\bibitem[Greene \latin{et~al.}(2019)Greene, Webber, Weare, and Berkelbach]{fast_randomized_iteration_berkelbach_2019}
Greene,~S.~M.; Webber,~R.~J.; Weare,~J.; Berkelbach,~T.~C. Beyond Walkers in Stochastic Quantum Chemistry: Reducing Error Using Fast Randomized Iteration. \emph{Journal of Chemical Theory and Computation} \textbf{2019}, \emph{15}, 4834--4850, PMID: 31390198\relax
\mciteBstWouldAddEndPuncttrue
\mciteSetBstMidEndSepPunct{\mcitedefaultmidpunct}
{\mcitedefaultendpunct}{\mcitedefaultseppunct}\relax
\EndOfBibitem
\bibitem[Greene \latin{et~al.}(2020)Greene, Webber, Weare, and Berkelbach]{improved_fri_berkelbach_2020}
Greene,~S.~M.; Webber,~R.~J.; Weare,~J.; Berkelbach,~T.~C. Improved Fast Randomized Iteration Approach to Full Configuration Interaction. \emph{Journal of Chemical Theory and Computation} \textbf{2020}, \emph{16}, 5572--5585, PMID: 32697909\relax
\mciteBstWouldAddEndPuncttrue
\mciteSetBstMidEndSepPunct{\mcitedefaultmidpunct}
{\mcitedefaultendpunct}{\mcitedefaultseppunct}\relax
\EndOfBibitem
\bibitem[Thom(2010)]{stochatsic_CC_thom_2010}
Thom,~A. J.~W. Stochastic Coupled Cluster Theory. \emph{Phys. Rev. Lett.} \textbf{2010}, \emph{105}, 263004\relax
\mciteBstWouldAddEndPuncttrue
\mciteSetBstMidEndSepPunct{\mcitedefaultmidpunct}
{\mcitedefaultendpunct}{\mcitedefaultseppunct}\relax
\EndOfBibitem
\bibitem[Filip \latin{et~al.}(2019)Filip, Scott, and Thom]{multiref_stochastic_cc_thom_2019}
Filip,~M.-A.; Scott,~C. J.~C.; Thom,~A. J.~W. Multireference Stochastic Coupled Cluster. \emph{Journal of Chemical Theory and Computation} \textbf{2019}, \emph{15}, 6625--6635, PMID: 31697497\relax
\mciteBstWouldAddEndPuncttrue
\mciteSetBstMidEndSepPunct{\mcitedefaultmidpunct}
{\mcitedefaultendpunct}{\mcitedefaultseppunct}\relax
\EndOfBibitem
\bibitem[Filip and Thom(2023)Filip, and Thom]{hybrid_ci_ccmc_thom_2023}
Filip,~M.-A.; Thom,~A. J.~W. {A hybrid stochastic configuration interaction–coupled cluster approach for multireference systems}. \emph{The Journal of Chemical Physics} \textbf{2023}, \emph{158}, 184101\relax
\mciteBstWouldAddEndPuncttrue
\mciteSetBstMidEndSepPunct{\mcitedefaultmidpunct}
{\mcitedefaultendpunct}{\mcitedefaultseppunct}\relax
\EndOfBibitem
\bibitem[Blunt \latin{et~al.}(2015)Blunt, Alavi, and Booth]{kryloc_projected_fciqmc_blunt_2015}
Blunt,~N.~S.; Alavi,~A.; Booth,~G.~H. Krylov-Projected Quantum Monte Carlo Method. \emph{Phys. Rev. Lett.} \textbf{2015}, \emph{115}, 050603\relax
\mciteBstWouldAddEndPuncttrue
\mciteSetBstMidEndSepPunct{\mcitedefaultmidpunct}
{\mcitedefaultendpunct}{\mcitedefaultseppunct}\relax
\EndOfBibitem
\bibitem[Neufeld and Thom(2020)Neufeld, and Thom]{quasinewton}
Neufeld,~V.~A.; Thom,~A. J.~W. Accelerating Convergence in Fock Space Quantum Monte Carlo Methods. \emph{Journal of Chemical Theory and Computation} \textbf{2020}, \emph{16}, 1503--1510\relax
\mciteBstWouldAddEndPuncttrue
\mciteSetBstMidEndSepPunct{\mcitedefaultmidpunct}
{\mcitedefaultendpunct}{\mcitedefaultseppunct}\relax
\EndOfBibitem
\bibitem[Blunt \latin{et~al.}(2015)Blunt, Smart, Kersten, Spencer, Booth, and Alavi]{semistochasticfciqmc}
Blunt,~N.~S.; Smart,~S.~D.; Kersten,~J. A.~F.; Spencer,~J.~S.; Booth,~G.~H.; Alavi,~A. Semi-stochastic full configuration interaction quantum Monte Carlo: Developments and application. \emph{The Journal of Chemical Physics} \textbf{2015}, \emph{142}, 184107\relax
\mciteBstWouldAddEndPuncttrue
\mciteSetBstMidEndSepPunct{\mcitedefaultmidpunct}
{\mcitedefaultendpunct}{\mcitedefaultseppunct}\relax
\EndOfBibitem
\bibitem[Saad(1992)]{saad1992}
Saad,~Y. \emph{Numerical Methods for Large Eigenvalue Problems}; Manchester University Press: Manchester, UK, 1992\relax
\mciteBstWouldAddEndPuncttrue
\mciteSetBstMidEndSepPunct{\mcitedefaultmidpunct}
{\mcitedefaultendpunct}{\mcitedefaultseppunct}\relax
\EndOfBibitem
\bibitem[Blunt \latin{et~al.}(2019)Blunt, Thom, and Scott]{fciqmc-preconditioning}
Blunt,~N.~S.; Thom,~A. J.~W.; Scott,~C. J.~C. Preconditioning and Perturbative Estimators in Full Configuration Interaction Quantum Monte Carlo. \emph{Journal of Chemical Theory and Computation} \textbf{2019}, \emph{15}, 3537--3551, PMID: 31050430\relax
\mciteBstWouldAddEndPuncttrue
\mciteSetBstMidEndSepPunct{\mcitedefaultmidpunct}
{\mcitedefaultendpunct}{\mcitedefaultseppunct}\relax
\EndOfBibitem
\bibitem[Azad and Buluc(2017)Azad, and Buluc]{spmspv}
Azad,~A.; Buluc,~A. A Work-Efficient Parallel Sparse Matrix-Sparse Vector Multiplication Algorithm. \emph{Proceedings - 2017 IEEE International Parallel and Distributed Processing Symposium (IPDPS)} \textbf{2017}, \emph{2017}\relax
\mciteBstWouldAddEndPuncttrue
\mciteSetBstMidEndSepPunct{\mcitedefaultmidpunct}
{\mcitedefaultendpunct}{\mcitedefaultseppunct}\relax
\EndOfBibitem
\bibitem[Lanczos(1950)]{Lanczos1950AnIM}
Lanczos,~C. An iteration method for the solution of the eigenvalue problem of linear differential and integral operators. \emph{Journal of research of the National Bureau of Standards} \textbf{1950}, \emph{45}, 255--282\relax
\mciteBstWouldAddEndPuncttrue
\mciteSetBstMidEndSepPunct{\mcitedefaultmidpunct}
{\mcitedefaultendpunct}{\mcitedefaultseppunct}\relax
\EndOfBibitem
\bibitem[Koch(2011)]{Koch2011}
Koch,~E. In \emph{The LDA+DMFT approach to strongly correlated materials, Forschungszentrum Jülich}; Pavarini,~E., Koch,~E., Vollhardt,~D., Lichtenstein,~A., Eds.; 2011\relax
\mciteBstWouldAddEndPuncttrue
\mciteSetBstMidEndSepPunct{\mcitedefaultmidpunct}
{\mcitedefaultendpunct}{\mcitedefaultseppunct}\relax
\EndOfBibitem
\bibitem[Saad(2003)]{YSaad}
Saad,~Y. \emph{Iterative methods for sparse linear systems}; {SIAM}, 2003\relax
\mciteBstWouldAddEndPuncttrue
\mciteSetBstMidEndSepPunct{\mcitedefaultmidpunct}
{\mcitedefaultendpunct}{\mcitedefaultseppunct}\relax
\EndOfBibitem
\bibitem[Bai \latin{et~al.}(2000)Bai, Demmel, Dongarra, Ruhe, and van~der Vorst]{templates}
Bai,~Z.; Demmel,~J.; Dongarra,~J.; Ruhe,~A.; van~der Vorst,~H. In \emph{Templates for the Solution of Algebraic Eigenvalue Problems}; Bai,~Z., Demmel,~J., Dongarra,~J., Ruhe,~A., van~der Vorst,~H., Eds.; Society for Industrial and Applied Mathematics, 2000\relax
\mciteBstWouldAddEndPuncttrue
\mciteSetBstMidEndSepPunct{\mcitedefaultmidpunct}
{\mcitedefaultendpunct}{\mcitedefaultseppunct}\relax
\EndOfBibitem
\bibitem[Dargel \latin{et~al.}(2012)Dargel, W\"ollert, Honecker, McCulloch, Schollw\"ock, and Pruschke]{spurious}
Dargel,~P.~E.; W\"ollert,~A.; Honecker,~A.; McCulloch,~I.~P.; Schollw\"ock,~U.; Pruschke,~T. Lanczos algorithm with matrix product states for dynamical correlation functions. \emph{Phys. Rev. B} \textbf{2012}, \emph{85}, 205119\relax
\mciteBstWouldAddEndPuncttrue
\mciteSetBstMidEndSepPunct{\mcitedefaultmidpunct}
{\mcitedefaultendpunct}{\mcitedefaultseppunct}\relax
\EndOfBibitem
\bibitem[Musser(1997)]{introsort}
Musser,~D.~R. Introspective sorting and selection algorithms. \emph{Software: Practice and Experience} \textbf{1997}, \emph{27}, 983--993\relax
\mciteBstWouldAddEndPuncttrue
\mciteSetBstMidEndSepPunct{\mcitedefaultmidpunct}
{\mcitedefaultendpunct}{\mcitedefaultseppunct}\relax
\EndOfBibitem
\bibitem[Werner \latin{et~al.}(2012)Werner, Knowles, Knizia, Manby, and Schütz]{molpro}
Werner,~H.-J.; Knowles,~P.~J.; Knizia,~G.; Manby,~F.~R.; Schütz,~M. Molpro: a general-purpose quantum chemistry program package. \emph{WIREs Computational Molecular Science} \textbf{2012}, \emph{2}, 242--253\relax
\mciteBstWouldAddEndPuncttrue
\mciteSetBstMidEndSepPunct{\mcitedefaultmidpunct}
{\mcitedefaultendpunct}{\mcitedefaultseppunct}\relax
\EndOfBibitem
\bibitem[rep()]{repo}
\url{https://github.com/wlj89/SFCI}\relax
\mciteBstWouldAddEndPuncttrue
\mciteSetBstMidEndSepPunct{\mcitedefaultmidpunct}
{\mcitedefaultendpunct}{\mcitedefaultseppunct}\relax
\EndOfBibitem
\bibitem[Chan \latin{et~al.}(2004)Chan, Kállay, and Gauss]{N2-fci-2004}
Chan,~G. K.-L.; Kállay,~M.; Gauss,~J. State-of-the-art density matrix renormalization group and coupled cluster theory studies of the nitrogen binding curve. \emph{The Journal of Chemical Physics} \textbf{2004}, \emph{121}, 6110--6116\relax
\mciteBstWouldAddEndPuncttrue
\mciteSetBstMidEndSepPunct{\mcitedefaultmidpunct}
{\mcitedefaultendpunct}{\mcitedefaultseppunct}\relax
\EndOfBibitem
\bibitem[Olivares-Amaya \latin{et~al.}(2015)Olivares-Amaya, Hu, Nakatani, Sharma, Yang, and Chan]{abinitiodmrg}
Olivares-Amaya,~R.; Hu,~W.; Nakatani,~N.; Sharma,~S.; Yang,~J.; Chan,~G. K.-L. The ab-initio density matrix renormalization group in practice. \emph{The Journal of Chemical Physics} \textbf{2015}, \emph{142}, 034102\relax
\mciteBstWouldAddEndPuncttrue
\mciteSetBstMidEndSepPunct{\mcitedefaultmidpunct}
{\mcitedefaultendpunct}{\mcitedefaultseppunct}\relax
\EndOfBibitem
\bibitem[Golub and Van~Loan(1996)Golub, and Van~Loan]{GoluVanl96}
Golub,~G.~H.; Van~Loan,~C.~F. \emph{Matrix Computations}, 3rd ed.; The Johns Hopkins University Press, 1996\relax
\mciteBstWouldAddEndPuncttrue
\mciteSetBstMidEndSepPunct{\mcitedefaultmidpunct}
{\mcitedefaultendpunct}{\mcitedefaultseppunct}\relax
\EndOfBibitem
\end{mcitethebibliography}

\end{document}